\documentclass[preprint,12pt]{aastex}

\shortauthors{Winn et al.~2008}
\shorttitle{Smaller Radius of XO-3b}

\begin{document}

%
\def\ltsima{$\; \buildrel < \over \sim \;$}
\def\lsim{\lower.5ex\hbox{\ltsima}}
\def\gtsima{$\; \buildrel > \over \sim \;$}
\def\gsim{\lower.5ex\hbox{\gtsima}}
\def\lam{\lambda=-1\fdg4 \pm 1\fdg1}

\def\jk{JK}
%

\bibliographystyle{apj}

\title{
The Transit Light Curve Project.\\
IX.~Evidence for a Smaller Radius of the Exoplanet XO-3b
}

\author{
Joshua N.\ Winn\altaffilmark{1},
Matthew J.\ Holman\altaffilmark{2},
Guillermo Torres\altaffilmark{2},
Peter McCullough\altaffilmark{3},
Christopher Johns-Krull\altaffilmark{4},
David W.\ Latham\altaffilmark{2},
Avi Shporer\altaffilmark{5},
Tsevi Mazeh\altaffilmark{5},
Enrique Garcia-Melendo\altaffilmark{6},
Cindy Foote\altaffilmark{7},
Gil Esquerdo\altaffilmark{2},
Mark Everett\altaffilmark{8}
}

\altaffiltext{1}{Department of Physics, and Kavli Institute for
  Astrophysics and Space Research, Massachusetts Institute of
  Technology, Cambridge, MA 02139, USA}

\altaffiltext{2}{Harvard-Smithsonian Center for Astrophysics, 60
  Garden Street, Cambridge, MA 02138, USA}

\altaffiltext{3}{Space Telescope Science Institute, 3700 San Martin
  Dr., Baltimore, MD 21218}

\altaffiltext{4}{Dept.\ of Physics and Astronomy, Rice University,
  6100 Main Street, MS-108, Houston, TX 77005}

\altaffiltext{5}{Wise Observatory, Raymond and Beverly Sackler Faculty
  of Exact Sciences, Tel Aviv University, Tel Aviv 69978, Israel}

\altaffiltext{6}{Esteve Duran Observatory, El Montanya, 08553 Seva,
  Barcelona, Spain}

\altaffiltext{7}{Vermillion Cliffs Observatory, 4175 E.~Red Cliffs
  Drive, Kanab, Utah, 84741}

\altaffiltext{8}{Planetary Science Institute, 1700 E.~Fort Lowell Rd.,
  Suite 106, Tucson, AZ 85719}

\begin{abstract}

  We present photometry of 13 transits of XO-3b, a massive transiting
  planet on an eccentric orbit. Previous data led to two inconsistent
  estimates of the planetary radius. Our data strongly favor the
  smaller radius, with increased precision: $R_p = 1.217 \pm
  0.073$~$R_{\rm Jup}$. A conflict remains between the mean stellar
  density determined from the light curve, and the stellar surface
  gravity determined from the shapes of spectral lines. We argue the
  light curve should take precedence, and revise the system parameters
  accordingly. The planetary radius is about $1\sigma$ larger than the
  theoretical radius for a hydrogen-helium planet of the given mass
  and insolation. To help in planning future observations, we provide
  refined transit and occultation ephemerides.

\end{abstract}


\section{Introduction}

The most intimate details about exoplanets have come from observations
of transits and occultations, as recently reviewed by Charbonneau et
al.~(2007a), Ksanfomality~(2007), and Seager~(2008). This is the ninth
publication of the Transit Light Curve (TLC) project, a series of
photometric investigations of transiting exoplanets. The short-term
goal of this project is the accurate determination of planetary radii
and other system parameters (Holman et al.~2007, Winn et al.~2007a),
the intermediate-term goal is detecting reflected light or thermal
emission with ground-based observations (Winn et al.~2008), and the
longer-term goal is seeking evidence for additional planets or
satellites in the pattern of measured transit times (Holman \& Murray
2005, Agol et al.~2005).

This paper is concerned with the determination of system parameters
for XO-3, which was discovered by Johns-Krull et al.~(2008; hereafter,
\jk), as part of the XO Project (McCullough et al.~2005). In this
system, a planet with a mass near the deuterium-burning limit of
13~M$_{\rm Jup}$ orbits an F5V star, with a period of 3.19~d and an
eccentricity of 0.26. The planet is the most massive transiting planet
yet reported. It is also one of only 4 transiting planets with an
obviously noncircular orbit. How such a massive planet formed, how it
achieved its tight orbit, and why the orbit is eccentric, are
interesting unanswered questions, and precise determinations of the
basic system parameters may help to answer them.

\jk\ found that a key parameter---the planetary radius---was
especially uncertain, with an allowed range from 1.17 to 2.10 times
the radius of Jupiter. This wide range encompassed the results of two
different methods for determining the planetary radius that gave
discrepant values. They found that if the true radius is near the low
end of this range, it can be accommodated by ordinary models for gas
giants of solar composition with the given mass and degree of
insolation, while if the radius is near the high end of the allowed
range, more complex and interesting models would require consideration
(see, e.g., Guillot \& Showman 2002, Bodenheimer et al.~2003, Chabrier
\& Baraffe 2007, Hansen \& Barman 2007, Burrows et al.~2007).

High-precision photometry of transits is one avenue for improving the
precision of the radius measurement. In \S~2, we describe our
observations and the production of the light curves. In \S~3, we
describe the procedure with which we estimated the system parameters
from the light curves. In \S~4, we present the results for the
planetary, stellar, and orbital parameters, as well as the transit
times and updated transit and occultation ephemerides, which are
useful for planning future observations.  In \S~5 we summarize, and
revisit the issue of the planetary radius and its theoretical
interpretation, in light of the new data.

\section{Observations and Data Reduction}

We observed XO-3 on 12 nights when transits were predicted to occur
according to the ephemeris of \jk. The observing dates and other
pertinent characteristics of the observations are given in Table~1.

On six of those nights, we used the 1.2m telescope at the Fred L.\
Whipple Observatory (FLWO) on Mt.\ Hopkins, Arizona. We used KeplerCam
(Szentgyorgi et al.~2005), which has a monolithic 4096$^2$ CCD
detector giving a $23\farcm1 \times 23\farcm 1$ field of view.  We
binned the images $2\times 2$, giving a scale of $0\farcs 68$ per
binned pixel. We used a Sloan $z$ filter, the reddest broad band
filter available, to minimize the effects of stellar limb darkening on
the transit light curves. On each night we attempted to observe as
much of the transit as possible, preferably starting at least 1~hr
prior to ingress and ending at least 1~hr after egress, although this
was not always possible. We defocused the telescope slightly to permit
exposure times of 10-15~s without saturating the brightest star in the
field. We also obtained dome-flat exposures and bias exposures for
calibration purposes.

On the other 6 nights, data were obtained with smaller telescopes. On
the night of 2008~Feb~10, we used the 0.5m telescope at Wise
Observatory, in Israel. We used a Santa Barbara Instrument Group
(SBIG) ST-10 XME CCD detector with $2148\times 1472$~pixels, giving a
field of view of $40\farcm 5 \times 27\farcm 3$ and a scale of
$1\farcs 1$ per pixel. No filter was used. More details about this
telescope and instrument are given by Brosch et al.~(2008). On
2007~Oct~9, we used two different telescopes at Vermillion Cliffs
Observatory, in Kanab, Utah: a 0.4m telescope with an SBIG ST-7e CCD
($1530\times 1020$~pixels, $0\farcs 87$~pixel$^{-1}$, $22\farcm 2
\times 14\farcm 8$~FOV), and a 0.6m telescope with an SBIG ST-8xe CCD
($765\times 510$~pixels, $1\farcs 31$~pixel$^{-1}$, $16\farcm 8 \times
11\farcm 2$~FOV). We used an $I$-band filter on the 0.4m telescope and
an $R$-band filter on the 0.6m telescope. Each telescope was used to
produce an independent light curve. On 2007~Sep~16, Oct~18, Oct~21,
and Nov~6, we used a 0.6m telescope with an SBIG ST-9XE CCD, located
at Esteve Duran Observatory in Seva, Spain. An $I$-band filter was
used for the first three of these events, and a $V$-band filter was
used for the Nov~06 event.

We used standard procedures for the overscan correction, trimming,
bias subtraction, and flat-field division. For the FLWO and Wise data,
we performed aperture photometry of XO-3 and 15-20 comparison
stars. The flux of XO-3 was divided by the sum of the fluxes of the
comparison stars, and then divided by a constant to give a unit mean
flux outside of transit. For the other data, only 2 comparison stars
were used. It was found in almost all cases that the out-of-transit
(OOT) flux was not a constant over the course of the night, perhaps
due to the effects of differential atmospheric extinction or slow
drifts in focus, pixel position, or other external variables. To
compensate, we solved for the ``OOT baseline function,'' a linear
function of time, as part of the fitting process described in the next
section. Figures~1 and 2 show the final light curves. Figure~3 is a
composite $z$-band light curve based on all of the FLWO data. Table~2
provides all of the data in numerical form.

\begin{figure}[p]
\epsscale{1.0}
\plotone{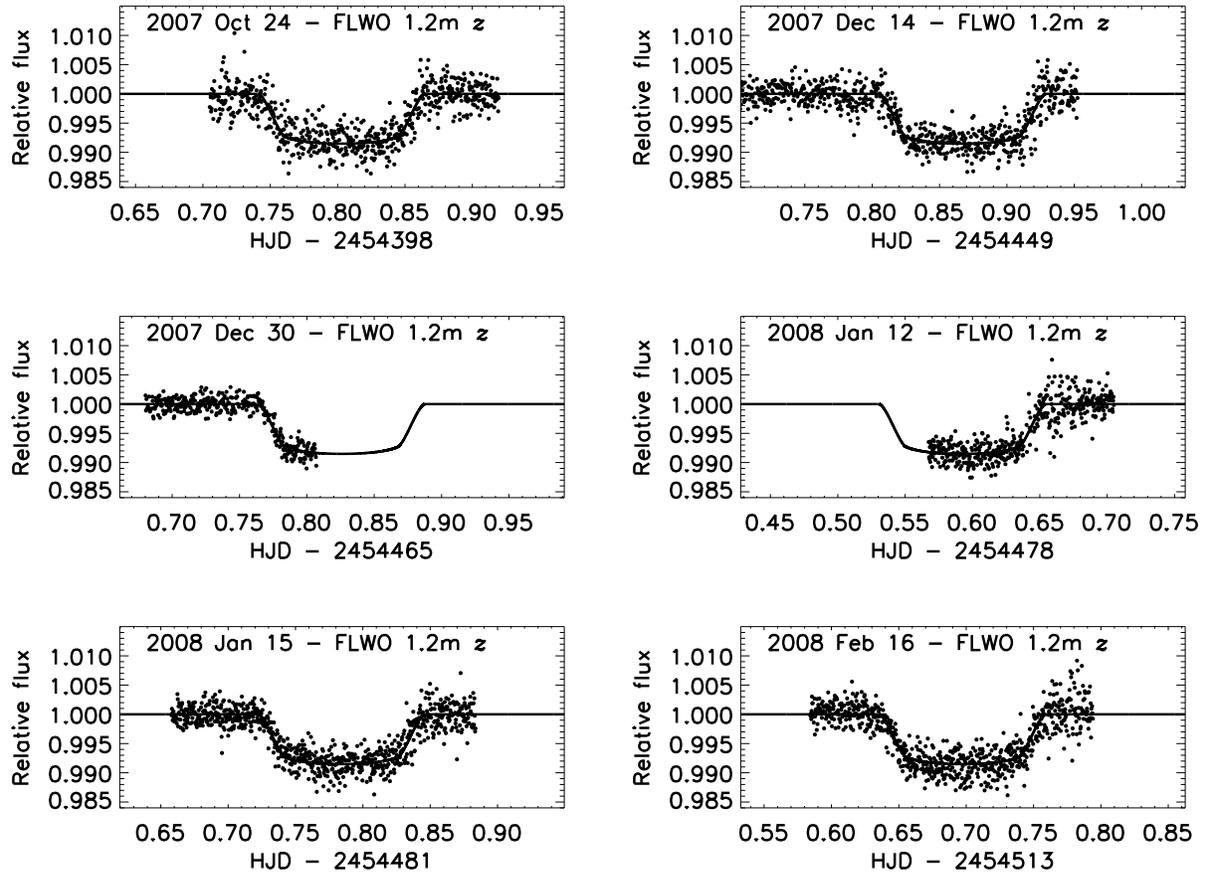}
\caption{ Relative $z$-band photometry of XO-3,
based on observations with the FLWO~1.2m telescope.
\label{fig:1}}
\end{figure}

\begin{figure}[p]
\epsscale{1.0}
\plotone{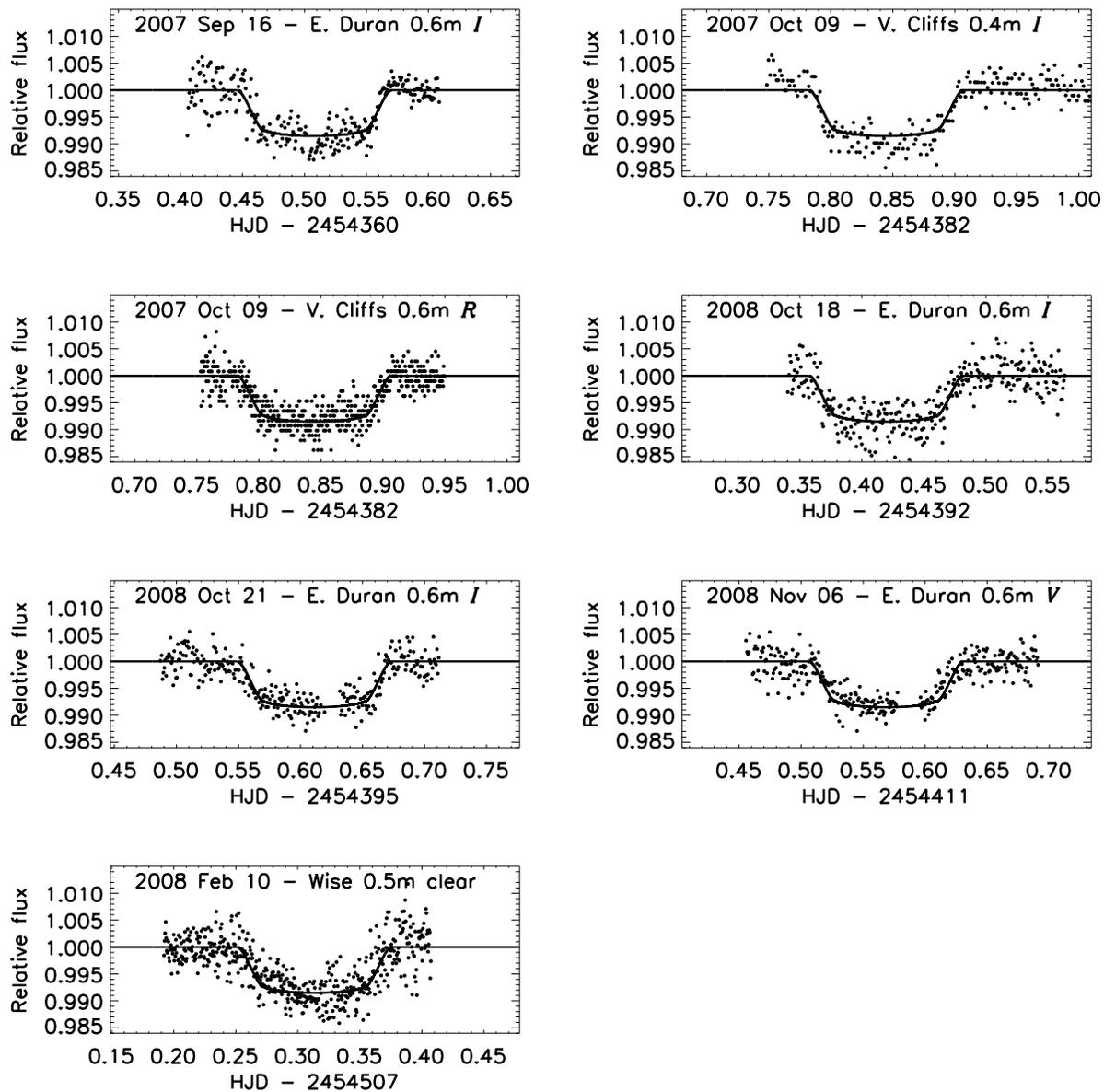}
\caption{ Relative photometry of XO-3,
based on observations with 0.4--0.6m telescopes.
\label{fig:2}}
\end{figure}

\begin{figure}[bp]
\epsscale{1.0}
\plotone{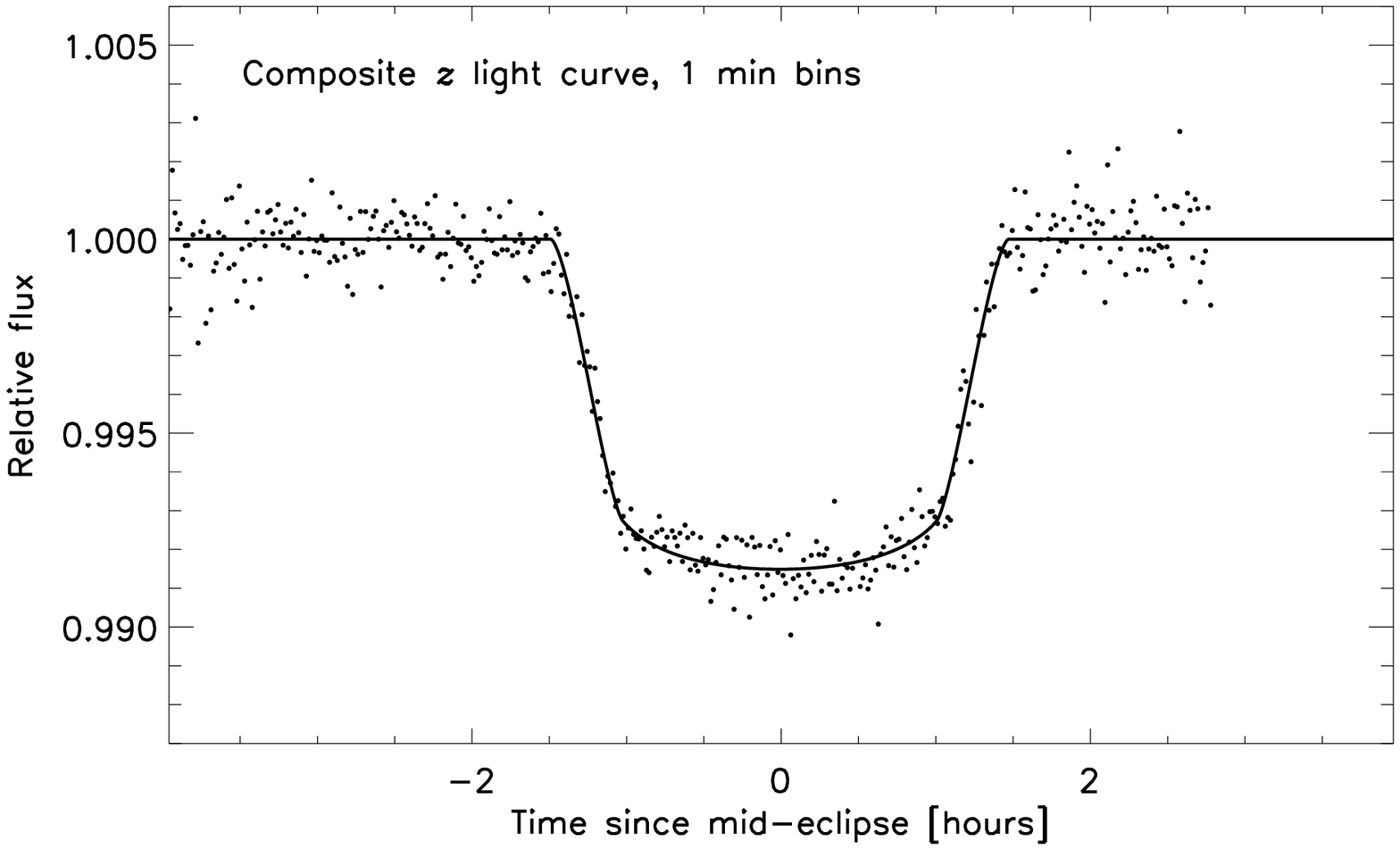}
\caption{ Composite $z$-band light curve of XO-3,
calculated by
subtracting the midtransit time from each time stamp,
and then averaging into 1~min bins.
\label{fig:3}}
\end{figure}

\section{Determination of System Parameters}

In order to determine the stellar, planetary, and orbital parameters,
we fitted a parametric model to the 13 photometric time series, as
well as the 21 radial velocity measurements of \jk. Our model and
fitting method were similar to those described in previous TLC papers
(see, e.g., Holman et al.~2006, Winn et al.~2007a). It is based on a
Keplerian orbit of two spherical bodies. The physical parameters were
the stellar mass and radius ($M_\star$ and $R_\star$); the planetary
mass and radius ($M_p$, $R_p$); the orbital period, inclination,
eccentricity, and argument of pericenter ($P$, $i$, $e$, $\omega$);
and a particular midtransit time ($T_c$). In addition, for each of the
two velocity data sets presented by \jk~(from the Hobby-Eberly
Telescope and the Harlan J.~Smith telescope), we allowed for an
arbitrary additive constant velocity. As mentioned in the previous
section, we also fitted for a linear function of time describing the
OOT flux (2 parameters per light curve).

A well-known degeneracy involves both of the bodies' masses and radii.
Only 3 of those 4 parameters can be determined independently. Three
parameters that {\it can}\, be determined independently are
$R_p/R_\star$, $M_\star/R_\star^3$ (Seager \& Mallen-Ornelas~2003),
and $M_p/R_p^2$ (Southworth et al.~2007). Rather than reparameterizing
in terms of those variables, we find it more convenient to fix the
stellar mass at some fiducial value and then use the scaling relations
$R_\star\propto M_\star^{1/3}$, $R_p\propto M_\star^{1/3}$, and
$M_p\propto M_\star^{2/3}$ as needed.

To calculate the relative flux as a function of the projected
separation of the planet and the star, we employed the analytic
formulas of Mandel \& Agol~(2002) to compute the integral of the
intensity over the unobscured portion of the stellar disk. We assumed
the limb darkening law to be quadratic. In some previous studies,
including our own, the limb-darkening coefficients have been fixed at
the tabulated values based on stellar-atmosphere models. A more
conservative approach is to fit for the limb-darkening law, since the
actual stellar brightness distribution is not known and may differ
from the tabulated limb-darkening law. However, it is generally not
possible to constrain more than one free parameter in the
limb-darkening law. Following the suggestion by Southworth~(2008), our
approach was to allow the linear coefficient ($a$) to vary freely, and
to fix the quadratic coefficient ($b$) at the appropriate value
tabulated by Claret~(2004). To determine the ``theoretical'' values of
the limb-darkening coefficients, we interpolated the ATLAS tables for
the stellar parameters $T_{\rm eff} = 6429$~K, $\log g = 4.244$~(cgs),
[Fe/H]$=-0.177$ and $v_t = 2.0$~km~s$^{-1}$. The interpolated values
are given in Table~4, as well as the results for the fitted linear
coefficient.

The fitting statistic was
\begin{equation}
\chi^2 =
\sum_{j=1}^{21}
\left[
\frac{v_j({\mathrm{obs}}) - v_j({\mathrm{calc}})}{\sigma_{v,j}}
\right]^2 + 
\sum_{j=1}^{6104}
\left[
\frac{f_j({\mathrm{obs}}) - f_j({\mathrm{calc}})}{\sigma_{f,j}}
\right]^2
,
\label{eq:chi2}
\end{equation}
where $v_j$(obs) is the radial velocity observed at time $j$,
$\sigma_{v,j}$ is the corresponding uncertainty, and $v_j$(calc) is
the calculated radial velocity. A similar notation applies to the
fluxes $f$. For the velocity uncertainties, we used the values
reported by \jk. For the flux uncertainties, we used a procedure that
attempts to account for time-correlated (``red'') noise, at least
approximately. For each of the 13 observed transits, we set the
uncertainty of each data point equal to the root-mean-squared (rms)
relative flux observed out of transit, multiplied by a factor
$\beta\geq 1$. The factor $\beta$ was determined using two different
methods (described at the end of this section), and the larger of the
two results was used in our final analysis.

We found the ``best fitting'' values of the model parameters, and
their uncertainties, using a Markov Chain Monte Carlo (MCMC) algorithm
[see Tegmark et al.~(2004) for applications to cosmological data, Ford
(2005) for radial-velocity data, Holman et al.~(2006) or Winn et
al.~(2007) for our particular implementation, and Burke et al.~(2007)
for a similar approach]. This algorithm creates a chain of points in
parameter space by iterating a jump function, which in our case was
the addition of a Gaussian random deviate to a randomly-selected
single parameter. If the new point has a lower $\chi^2$ than the
previous point, the jump is executed; if not, the jump is executed
with probability $\exp(-\Delta\chi^2/2)$. We set the sizes of the
random deviates such that $\sim$40\% of jumps are executed. We create
a number of chains from different starting conditions to verify they
all converge to the same basin in parameter space, and then we merge
them for our final results. The phase-space density of points in the
chain is an estimate of the joint {\it a posteriori}\, probability
distribution of all the parameters, from which may be calculated the
probability distribution for an individual parameter by marginalizing
over all of the others.

The fitting procedure had 4 basic steps. First, we performed a joint
fit of all the light curves along with the radial velocities, to
determine provisional values of the orbital period and physical
parameters. Second, we measured individual midtransit times, by
performing a MCMC analysis of each light curve with only three free
parameters: the zero point and slope of the linear function describing
the out-of-transit flux, and the midtransit time. We fixed $R_p$,
$R_\star$, and $i$ at the best-fitting values determined from the
ensemble. There is no need to fit the midtransit times simultaneously
with $\{R_p, R_\star, i\}$ because the errors in those parameters are
uncorrelated with the error in the midtransit time. Third, we
recomputed the transit ephemeris using the newly measured midtransit
times (see \S~\ref{subsec:ephemeris} for more details on this step).
Fourth, we fixed the orbital period and midtransit times at the values
just determined, and performed another joint fit of all the
radial-velocity and photometric data, to obtain final estimates of the
model parameters and their uncertainties. The results from this final
computation did not differ significantly from the results of the
initial joint fit.

As mentioned previously, we used two different methods to estimate the
factor $\beta$ by which time-correlated noise effectively increases
the flux uncertainties. We refer to the first method as the
``time-averaging'' method, which has been described previously by Winn
et al.~(2007b) and is closely related to a method used by Gillon et
al.~(2006). For each light curve we found the best-fitting model and
calculated $\sigma_1$, the standard deviation of the unbinned
residuals between the observed and calculated fluxes. Next we averaged
the residuals into $M$ bins of $N$ points and calculated the standard
deviation $\sigma_N$ of the binned residuals. In the absence of red
noise, we would have expected\footnote{We thank G.~Kovacs for pointing
  out that we erroneously neglected the factor $\sqrt{M/(M-1)}$ in
  previous analyses. Typically $M>5$ and this factor is smaller than
  1.12.}
\begin{equation}
\sigma_N = \frac{\sigma_1}{\sqrt{N}} \sqrt{\frac{M}{M-1}},
\end{equation}
but often $\sigma_N$ is larger than this by a factor $\beta$. We found
that $\beta$ depends only weakly on the choice of averaging time
$\tau$, generally rising to an asymptotic value at $\tau\approx
10$~min. We denote by $\beta_1$ the median of these factors when using
averaging times ranging from 15-30~min (the approximate duration of
ingress or egress).

We refer to the second method as the ``rosary-bead'' method, which has
been used previously by many investigators (e.g., Bouchy et al.~2005,
Southworth~2008). For each light curve, we found the best-fitting
model and computed the time series of $N$ residuals. We then added
these residuals to the model light curve after time-shifting them by
$M$ samples with a periodic boundary condition, i.e., the $i$th
residual was assigned to the time stamp $(i+M)$~mod~$N$. We repeated
this for all $M<N$, then fitted each of these synthetic light curves
and took the standard deviation of the results to be the error
estimates for the parameters. This is a variant of the well-known
bootstrap technique that preserves the temporal correlations among the
residuals. We denote by $\beta_2$ the ratio between the error estimate
returned by the rosary-bead method, and the error estimate assuming
uncorrelated errrors.

In general $\beta_1$ and $\beta_2$ are specific to each parameter of
the model, but for simplicity we assumed they are the same for all
parameters, and to calculate them we focused on the determination of
midtransit times. For each of the 13 light curves we compared the
error bar in $T_c$ as obtained through the time-averaging method, and
as obtained with the rosary-bead method.  For the 13 light curves,
$\beta_2 / \beta_1$ varied from 0.86 to 1.47, with a mean of 1.13 and
a standard deviation of 0.21. Thus the two methods gave similar
results, and the rosary-bead method tended to produce larger error
estimates. For our final results, we assigned each light curve the
value $\beta = \max(\beta_1,\beta_2)$. These choices of $\beta$ are
given in Table~1. All of these procedures may be fairly criticized for
lacking statistical rigor, but experience has shown that the more
common procedure of setting $\sigma_{f,i} = \sigma_1$ results in
underestimated uncertainties in the model parameters, as demonstrated
by a lack of agreement between the results of different but presumably
equivalent data sets.

\section{Results}

Table~1 gives all of the newly measured transit times. Table~3 gives
the results for the planetary, stellar, and orbital parameters, as
well as many other quantities of intrinsic interest or importance for
planning follow-up observations. As an example of the latter, the
quantity $(R_p/a)^2$ is the planet-to-star flux ratio at opposition,
for a geometric albedo of unity, and as such it is relevant to
pursuing observations of reflected light from the planet. Another
example is the amplitude of the Rossiter-McLaughlin effect, given by
$(R_p/R_\star)^2 (v\sin i_\star)$, where $v\sin i_\star$ is the
projected rotation rate of the star (see, e.g., Gaudi \& Winn
2007). The labels A--E, explained in the table caption, are an attempt
to clarify which quantities are determined independently from our
analysis, which quantities are functions of those independent
parameters, and which quantities depend on our isochrone analysis to
break the fitting degeneracy between the stellar mass and radius.
Table~4 gives the results for the limb-darkening coefficients.

\subsection{The stellar and planetary radii}

As discussed in the previous section, the joint analysis of the light
curves and velocities cannot independently determine the masses and
radii of both bodies. Some external information about the star or the
planet must be introduced to break the fitting degeneracies
$M_p\propto M_\star^{2/3}$ and $R_p \propto R_\star \propto
M_\star^{1/3}$. Our approach was to seek consistency between the
observed spectroscopic properties of the star, the stellar mean
density that is derived from the transit light curves, and theoretical
models of stellar evolution. This is the same approach (and uses the
same software) that was described in detail by Sozzetti et al.~(2007)
and Torres et al.~(2008). Here we summarize the procedure, and refer
the reader to those papers for more details.

There were two sets of inputs. First, we used the values of the
effective temperature $T_{\rm eff}$, surface gravity $\log g$, and
metallicity [Fe/H] of the host star, as reported by \jk, based on a
parametric fit to the optical spectrum of XO-3 using the {\it
  Spectroscopy Made Easy}\, (SME) program (Valenti \& Piskunov 1996,
Valenti \& Fischer 2005). Second, we used the scaled semimajor axis
$a/R_\star$ and the orbital parameters from our joint analysis of the
photometry and radial-velocity data. Given $a/R_\star$ and the orbital
parameters, it is possible to derive the mean stellar density, using
Kepler's Law (Seager \& Mallen-Ornelas~2003). We use the symbol
$\rho_\star$ to refer to the mean density determined in this fashion.

We used the Yonsei-Yale (Y$^2$) stellar evolution models by Yi et
al.~(2001) and Demarque et al.~(2004). We computed isochrones for the
full range of metallicities allowed by the data, and for stellar ages
ranging from 0.1 to 14~Gyr, seeking points that gave agreement with
the observed $T_{\rm eff}$ and one of the two gravity indicators
($\log g$ and $\rho_\star$). For each stellar property (mass, radius,
and age), we took a weighted average of the points on each
isochrone. The weights were based on the agreement with the observed
temperature, metallicity, and gravity indicator, and a factor taking
into account the number density of stars along each isochrone
(assuming a Salpeter mass function).

In almost all of the 23 cases examined by Torres et al.~(2008), the
results when using either $\log g$ or $\rho_\star$ as the gravity
indicator were in agreement, and greater precision was obtained with
$\rho_\star$, often by a factor of 2 or more. Torres et al.~(2008) did
not consider the case of XO-3, but using the same technique we find
poor agreement between the results of using the two independent
gravity indicators. Using $\rho_\star$ results in a less massive and
smaller star, with a higher mean density and a stronger surface
gravity. This in turn gives a less massive and smaller planet. One way
to frame the discrepancy is that by using $\rho_\star$ as the gravity
indicator, the isochrone analysis gives a stellar surface gravity of
$(\log g)_{\rm phot} = 4.244\pm 0.041$, as compared to the SME-derived
value of $\log g = 3.950\pm 0.062$. The difference is $0.294 \pm
0.074$, which is inconsistent with zero at the 4$\sigma$
level. Clearly something is amiss with either our interpretation of
the light curve, or the SME determination of $\log g$, or both. The
same conflict was already apparent between the light-curve analysis
and the isochrone analysis of \jk. We have improved the precision of
the light-curve parameters by factors of 3 or more, and the conflict
with the spectroscopic determination of $\log g$ has been sharpened.

Some further thoughts on the tension between $\rho_\star$ and $\log g$
are given in \S~5. For the results given in Table~3, we proceeded
under the assumption that the error in the spectroscopic determination
of $\log g$ was greatly underestimated. We disregarded the
spectroscopic $\log g$ while performing the isochrone analysis, and we
also increased the error bars on $T_{\rm eff}$ and [Fe/H], since the
errors in those three quantities are highly correlated when fitting
models to the features observed in optical spectra. We increased the
error in $T_{\rm eff}$ from 50~K to 100~K, and in [Fe/H] from
0.023~dex to 0.08~dex, which we believe to be conservative choices,
and are consistent with similar judgments made by Torres et al.~(2008)
in their homogeneous analysis of transiting systems.

\subsection{The transit and occultation ephemerides}
\label{subsec:ephemeris}

We calculated a photometric ephemeris for the transits of XO-3 using
the 13 midtransit times given in Table~1 and the 16 midtransit times
measured by the XO Extended Team and reported previously by \jk. We
fitted a linear function of transit epoch $E$,
\begin{equation}
T_c(E) = T_c(0) + E P.
\label{eq:ephemeris}
\end{equation}
The fit had $\chi^2=29.6$ with 27 degrees of freedom, or $\chi^2/N_{\rm
  dof} = 1.10$. The results were $T_c(0) = 2454449.86816 \pm
0.00023$~[HJD] and $P = 3.1915239\pm 0.0000068$~days.  Our derived
period agrees with the value $3.19154\pm 0.00014$~days determined by
\jk\, and is about 20 times more precise. Figure~4 is the O$-$C
(observed minus calculated) diagram for the transit times. In this
calculation, we did not use the 4 midtransit times that were based on
data from the XO survey instrument, because those data had
unquantified and apparently large uncertainties. Nevertheless, all of
the observed times are plotted in Figure~3, and are seen to be at
least roughly consistent with the new ephemeris.

To help in planning observations of occultations (secondary eclipses)
of XO-3b, we have also used our model results to predict the timing,
duration, and impact parameter of the occultations. Because of the
eccentric orbit, occultations and transits are not separated by
exactly one-half of the orbital period, and do not have the same
duration or impact parameter. Based on our model of the system, we
expect occultations to occur $2.109\pm 0.034$~days after transits. The
predicted occultation ephemeris is given in Table~3.

\begin{figure}[ht]
\epsscale{1.0}
\plotone{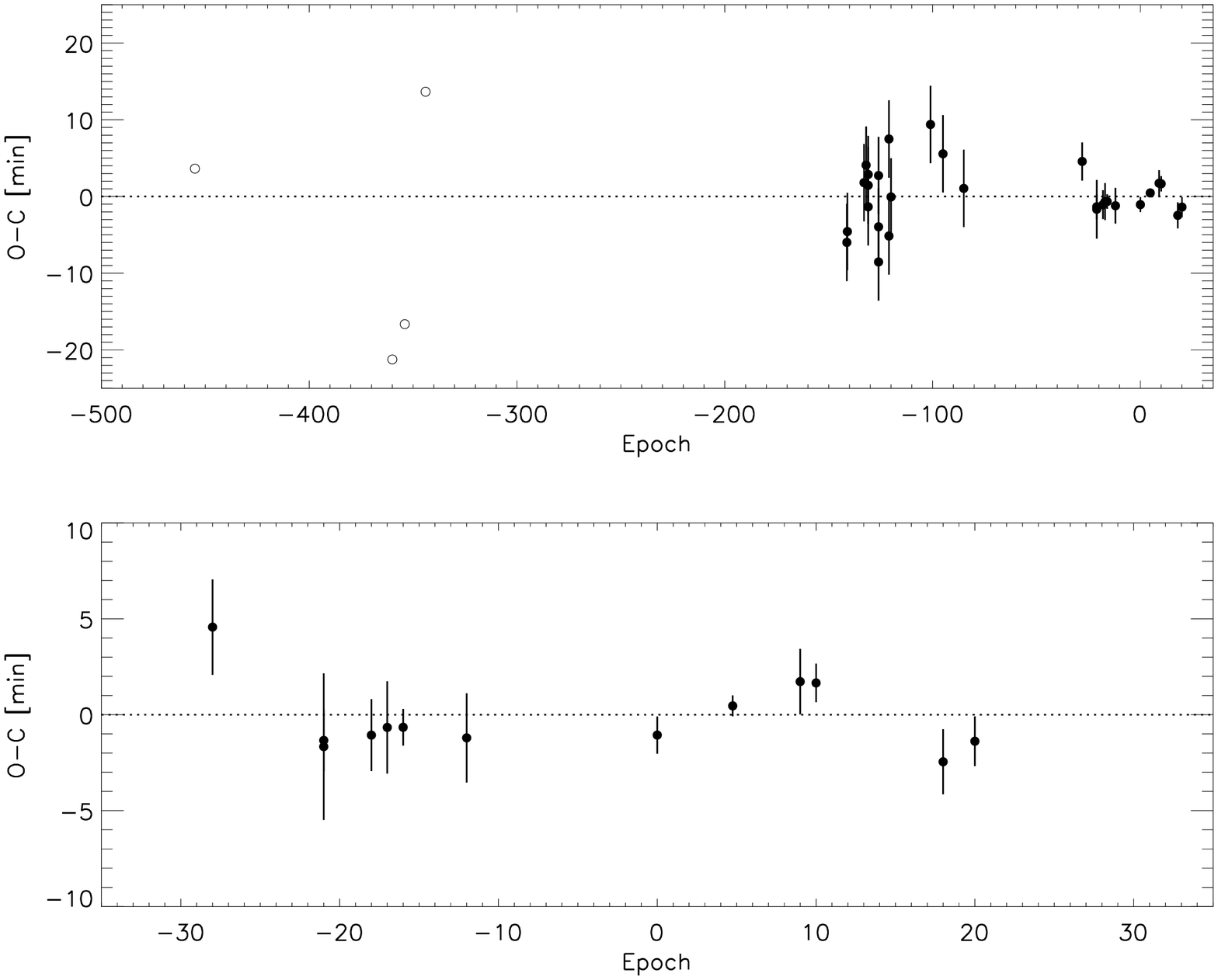}
\caption{Transit timing residuals for XO-3b. The calculated times,
  using the ephemeris derived in \S~4.2, have been subtracted from the
  observed times. The filled symbols are the data from
  this work and from the XO Extended Team observations reported by
  \jk. Those data were used to calculate the transit ephemeris.
  The unfilled symbols are the data from the XO Survey instruments,
  which were not used to calculate the transit ephemeris.
  \label{fig:4}}
\end{figure}

\section{Summary and Discussion}

We have presented new photometry spanning transits of the exoplanet
XO-3. The photometry greatly improves the precision with which the
light-curve parameters are known. In particular, the planet-to-star
radius ratio is known to within 0.6\%, an improvement by a factor of
6. The inclination angle is now known to within 0.54~deg, an
improvement by a factor of 2.5. A third light-curve parameter, the
scaled semimajor axis ($a/R_\star$), has also been refined by a factor
of a few, and was used (along with the orbital period and Kepler's
Law) to calculate $\rho_\star$, the stellar mean density. We found
that the photometric result for $\rho_\star$ is incompatible (at the
4$\sigma$ level) with the previous spectroscopic determination of
$\log g$, in the sense that theoretical stellar-evolution models
cannot accommodate both values along with the observed effective
temperature and metallicity of the star.

Because of this conflict, it is worth reviewing how $\rho_\star$ and
$\log g$ were determined. The photometric determination of
$\rho_\star$ is based on a fit to the light curve with 3 relevant free
parameters, and the application of Kepler's Law (Seager \&
Mallen-Ornelas 2003). The spectroscopic determination of $\log g$ is
based on the interpretation of pressure-sensitive features of the
stellar spectrum, especially the widths of the wings of selected
absorption lines. The interpretation is performed by comparison to
theoretical models of stellar atmospheres. For XO-3 this comparison
was performed with SME (Valenti \& Piskunov~1996, Valenti \&
Fischer~2005), an automated analysis program that fits a model to an
optical spectrum by adjusting many free parameters, of which the most
relevant are the effective temperature, surface gravity, projected
rotation rate, and metal abundances. The model is based on
plane-parallel stellar atmosphere models in local thermodynamic
equilibrium, and reasonable assumptions regarding instrumental
broadening and turbulent broadening mechanisms. Empirical corrections
are applied to the parameters based on an SME analysis of the Solar
spectrum.

The spectroscopic method for determining $\log g$ is more complex than
the photometric method for determining $\rho_\star$. In addition, it
is important to recognize that the quoted error in the spectroscopic
determination of \jk\ ($\log g = 3.950\pm 0.062$) represents the
standard error of the mean of the results of fitting 10 independent
spectra of XO-3. Thus, the error bar refers to the repeatability or
precision of the result, and not its accuracy. Valenti \&
Fischer~(2005) assessed the accuracy of SME by comparing two methods
of determining surface gravity: (1) the purely spectroscopic method
described in the previous paragraph; (2) the surface gravity that
follows from the observed stellar luminosity (for stars with measured
parallaxes), effective temperature, and metallicity, by requiring
consistency with theoretical isochrones of stellar-evolution
models. They found a systematic offset of 0.1~dex and a large scatter
(see \S~7.4 of that work). \jk\, repeated this comparison for stars
with similar temperatures to XO-3, finding a scatter of about 0.1~dex
and some cases in which the discrepancy is $\approx$0.3~dex. Valenti
\& Fischer~(2005) also compared the SME results for $\log g$ with
spectroscopic results that have been obtained by other authors,
finding a scatter of about 0.15~dex, and discrepancies as large as
0.3~dex. These general comparisons cannot speak to the specific case
of XO-3, but they suggest that the 4$\sigma$ discrepancy between the
spectroscopic and photometric methods in the present study is not as
serious as it may seem. The true error in the spectroscopic
determination of $\log g$ is probably larger than 0.062.

An upward revision of the stellar $\log g$ corresponds to a downward
revision of the planetary radius, to $1.217\pm 0.073$~$R_{\rm Jup}$.
How does this result compare to the radius that is expected on
theoretical grounds? Fortney et al.~(2007) have computed models for
planets over a wide range of masses, compositions, ages, and
irradiation levels, and provided the results in a convenient tabular
form. Interpolation of those tables for a coreless, pure
hydrogen-helium planet with properties appropriate for the XO-3 system
($M_p = 11.8$~$M_{\rm Jup}$, $a=0.045$~AU, $L_\star = 2.9$~$L_\odot$,
age~2.82~Gyr) gives a theoretical radius of 1.14~$R_{\rm Jup}$.

Adding as much as $\sim$100~$M_\oplus$ of heavy elements would
decrease the theoretical radius by a few per cent. On the other hand,
the models of Fortney et al.~(2008) define the planetary surface as
the 1~bar pressure level, whereas a much lower pressure is appropriate
for comparison to transit observations. This ``transit radius effect''
will increase the theoretical radius by a few per cent (Burrows et
al.~2007). Assuming the combination of these effects to be small, the
observed radius is 1$\sigma$ larger than the theoretical radius. Thus,
the photometric analysis leads to a planetary radius that is only a
little larger than the models of Fortney et al.~(2007) would predict,
similar to the case of HAT-P-1b (Winn et al.~2007b), and not nearly as
``inflated'' as some other examples in the literature such as
HD~209458b (Charbonneau et al.~2000, Henry et al.~2000), WASP-1b
(Collier Cameron et al.~2007, Charbonneau et al.~2007b), and TrES-4b
(Mandushev et al.~2007).

Although we have argued that the photometric determination of
$\rho_\star$ should take precedence over the spectroscopic
determination of $\log g$, it would be more definitive to settle the
issue by measuring the trigonometric parallax of XO-3, as suggested by
\jk. Our photometric analysis predicts that the distance will be found
to be $174\pm 18$~pc, based on the stellar luminosity inferred from
theoretical isochrones, the $V$ magnitude of $9.80\pm 0.03$, and the
assumption of negligible extinction. The spectrosopic analysis of
\jk~predicted a greater distance, $260\pm 23$~pc. An interesting but
more challenging prospect for determining the stellar mass (and hence
its radius) is to measure the general-relativistic periastron
precession of $2\arcmin$~yr$^{-1}$ by precise long-term timing of
transits and occultations (Heyl \& Gladman 2007).

In addition to pinning down the correct value of the radius, there are
other reasons to pursue further observations of XO-3, of which many
are related to its sizable orbital eccentricity. One consequence of
the eccentricity is that the planet experiences significant variations
in stellar insolation over the 3.2~d orbital period. The time-variable
response of the planet's atmosphere may be detectable through
mid-infrared photometry (Langton \& Laughlin 2008). Whatever mechanism
produced the large eccentricity may also have produced a large
inclination angle relative to the stellar equatorial plane, an angle
that can be measured through observations of the Rossiter-McLaughlin
effect. Narita et al.~(2007) have presented this type of evidence for
a significant orbital tilt in the HD~17156 system.

If, on the other hand, the stellar rotation axis is well-aligned with
the orbital axis, then the combination of the measurements of $v\sin
i$, $i$, and $R_\star$ give a stellar rotation period of $P_{\rm rot}
= 3.73\pm 0.23$~days. This is not too far from the orbital period of
3.19~days, suggesting that spin-orbit interactions may be unusually
strong, perhaps even strong enough to excite the orbital eccentricity
to the observed value.

Another way to produce an eccentricity is through stable long-term
gravitational interactions with another planet. Although the
midtransit times we have recorded are nearly consistent with a
constant period, and hence do not provide {\it prima facie}\, evidence
for any additional bodies in the system, we have achieved a precision
of 1-2~min using relatively small telescopes. After a few more
seasons, a pattern may yet emerge.

\acknowledgments We thank the anonymous referee for several insightful
suggestions. We are grateful for partial support for this work from
NASA Origins grants NNG06GH69G (to M.J.H.), NNG04LG89G (to G.T.),
05-SSO05-86 (to C.M.J.-K.), and NAG5-13130 (to P.R.M.). This work was
also partly supported by Grant no.~2006234 from the United
States--Israel Binational Science Foundation (BSF), Jerusalem,
Israel. KeplerCam was developed with partial support from the Kepler
Mission under NASA Cooperative Agreement NCC2-1390 and the Keplercam
observations described in this paper were partly supported by grants
from the Kepler Mission to SAO and PSI.

\begin{deluxetable}{lccccccc}
\tabletypesize{\scriptsize}
\tablecaption{Journal of Observations of XO-3\label{tbl:journal}}
\tablewidth{0pt}

\tablehead{
\colhead{Date} &
\colhead{Telescope} &
\colhead{Filter} &
\colhead{Cadence} &
\colhead{RMS} &
\colhead{Red noise factor} &
\colhead{Effective noise} &
\colhead{Midtransit time} \\
\colhead{[UT]} &
\colhead{} &
\colhead{} &
\colhead{$\Gamma$~[min$^{-1}$]} &
\colhead{$\sigma$} &
\colhead{$\beta$} &
\colhead{$\sigma\beta/\sqrt{\Gamma}$} &
\colhead{[HJD]}
}

\startdata
 2007~Sep~16 & E.~Duran 0.6m  & $I$ & $0.96$ & $0.0024$ & $2.02$ & $0.0049$ & $2454360.50866 \pm 0.00173$ \\  
 2007~Oct~09 & V.~Cliffs 0.4m & $I$ & $0.54$ & $0.0023$ & $3.06$ & $0.0096$ & $2454382.84500 \pm 0.00265$ \\  
 2007~Oct~09 & V.~Cliffs 0.6m & $R$ & $1.82$ & $0.0022$ & $1.58$ & $0.0026$ & $2454382.84523 \pm 0.00112$ \\  
 2007~Oct~18 & E.~Duran 0.6m  & $I$ & $0.96$ & $0.0027$ & $1.18$ & $0.0033$ & $2454392.41999 \pm 0.00130$ \\  
 2007~Oct~21 & E.~Duran 0.6m  & $I$ & $0.96$ & $0.0022$ & $1.93$ & $0.0043$ & $2454395.61179 \pm 0.00167$ \\  
 2007~Oct~24 & FLWO 1.2m      & $z$ & $2.08$ & $0.0023$ & $1.10$ & $0.0017$ & $2454398.80332 \pm 0.00066$ \\  
 2007~Nov~06 & E.~Duran 0.6m  & $V$ & $0.96$ & $0.0021$ & $1.42$ & $0.0030$ & $2454411.56904 \pm 0.00161$ \\  
 2007~Dec~14 & FLWO 1.2m      & $z$ & $2.08$ & $0.0020$ & $1.63$ & $0.0023$ & $2454449.86742 \pm 0.00067$ \\  
 2007~Dec~30 & FLWO 1.2m      & $z$ & $2.08$ & $0.0011$ & $1.00$ & $0.0008$ & $2454465.82610 \pm 0.00038$\tablenotemark{a} \\  
 2008~Jan~12 & FLWO 1.2m      & $z$ & $2.50$ & $0.0020$ & $1.31$ & $0.0017$ & $2454478.59308 \pm 0.00119$\tablenotemark{a} \\  
 2008~Jan~15 & FLWO 1.2m      & $z$ & $2.50$ & $0.0018$ & $1.57$ & $0.0018$ & $2454481.78455 \pm 0.00070$ \\  
 2008~Feb~10 & Wise 0.5m      & none &$1.61$ & $0.0030$ & $1.18$ & $0.0028$ & $2454507.31319 \pm 0.00118$ \\  
 2008~Feb~16 & FLWO 1.2m      & $z$ & $2.50$ & $0.0022$ & $1.69$ & $0.0024$ & $2454513.69768 \pm 0.00090$     
\enddata

\tablenotetext{a}{Only a partial transit was observed.}

\tablecomments{Column 1 gives the UT date at the start of the night.
  Column 4 gives $\Gamma$, the median number of data points per
  minute. Column 5 gives $\sigma$, the root-mean-squared (RMS)
  relative flux after subtracting the best-fitting model. Column 6
  gives the scaling factor $\beta$ that was applied to the
  single-point flux uncertainties to account for red noise (see
  \S~3). Column 7 gives the effective noise per minute, defined as
  $\sigma\beta/\sqrt{\Gamma}$.}

\end{deluxetable}

\begin{deluxetable}{lcccc}
\tabletypesize{\normalsize}
\tablecaption{Photometry of XO-3\label{tbl:photometry}}
\tablewidth{0pt}

\tablehead{
\colhead{Observatory Code\tablenotemark{a}} &
\colhead{Filter} &
\colhead{Heliocentric Julian Date} & 
\colhead{Relative flux}
}

\startdata
  $  1$  &       z  &  $  2454398.70513$  &  $         0.9977$ \\
  $  1$  &       z  &  $  2454398.70546$  &  $         1.0000$ \\
  $  1$  &       z  &  $  2454398.70578$  &  $         0.9994$ \\
  $  1$  &       z  &  $  2454398.70611$  &  $         0.9970$ \\
  $  1$  &       z  &  $  2454398.70644$  &  $         0.9975$ \\
  $  1$  &       z  &  $  2454398.70676$  &  $         1.0008$
\enddata 

\tablenotetext{a}{
(1) Fred L.~Whipple Observatory 1.2m telescope, Arizona, USA.
(2) Wise Observatory 0.5m telescope, Israel.
(3) Vermillion Cliffs Observatory 0.4m telescope, Utah, USA.
(4) Esteve Duran Observatory 0.6m telescope, Seva, Spain.
}

\tablecomments{The time stamps represent the Heliocentric Julian Date
  at the time of mid-exposure. We intend for this Table to appear in
  entirety in the electronic version of the journal. An excerpt is
  shown here to illustrate its format. The data are also available
  from the authors upon request.}

\end{deluxetable}

\begin{deluxetable}{lccc}
\tabletypesize{\scriptsize}
\tablecaption{System Parameters of XO-3\label{tbl:params}}
\tablewidth{0pt}

\tablehead{
\colhead{Parameter} & \colhead{Value} & \colhead{68.3\% Conf.~Limits} & \colhead{Comment}
}

\startdata
{\it Transit parameters:} & & & \\
Orbital period, $P$~[d]                             & $3.1915239$     & $\pm 0.0000068$  & A  \\
Midtransit time~[HJD]                               & $2454449.86816$ & $\pm 0.00023$    & A  \\
Planet-to-star radius ratio, $R_p/R_\star$            & $0.09057$      &  $\pm 0.00057$   & A  \\
Orbital inclination, $i$~[deg]                      &  $84.20$        &  $\pm 0.54$      & A  \\
Scaled semimajor axis, $a/R_\star$                    &  $7.07$        &  $\pm 0.31$      & A  \\
Transit impact parameter                            &   $0.705$       &  $\pm 0.023$     & B  \\
Transit duration~[hr]                               &   $2.989$       &  $\pm 0.029$     & B  \\
Transit ingress or egress duration~[hr]             &   $0.466$       &  $\pm 0.033$     & B  \\
RM figure of merit, $(v\sin i_\star)(R_p/R_\star)^2$~[m~s$^{-1}$] & $152.2$ & $\pm 2.3$      & B,C \\  
& & & \\
{\it Occultation parameters (predicted):} & & & \\
Midoccultation time~[HJD]                           & $2454451.977$ &  $\pm 0.034$ & B  \\
Occultation duration~[hr]                           &  $2.86$  &  $-0.014$,$+0.080$ & B  \\
Occultation ingress or egress duration~[hr]         &  $0.353$ &  $-0.027$,$+0.067$ & B  \\
Occultation impact parameter                        &  $0.614$ &  $\pm 0.050$  & B  \\
Reflected-light figure of merit, $(R_p/a)^2$         &  $0.000164$  & $\pm 0.000015$ & B  \\
& & & \\
{\it Other orbital parameters:} & & & \\
Orbital eccentricity, $e$                           & $0.260$   &  $\pm 0.017$    & A  \\
Argument of pericenter, $\omega$~[deg]              & $345.8$   &  $\pm 7.3$      & A  \\
Velocity semiamplitude, $K$~[m~s$^{-1}$]             & $1463$    &  $\pm 53$       & A  \\
Planet-to-star mass ratio, $M_p/M_\star$             &  $0.00927$  & $\pm 0.00036$  & E  \\
Semimajor axis~[AU]                                 &  $0.0454$   &  $\pm 0.00082$ & E  \\
& & & \\
{\it Stellar parameters:} & & & \\
Mass, $M_\star$~[M$_{\odot}$]                  &  $1.213$     &  $\pm 0.066$  & E  \\
Radius, $R_\star$~[R$_{\odot}$]                &  $1.377$     &  $\pm 0.083$  & E  \\
Mean density, $\rho_\star$~[g~cm$^{-3}$]      &  $0.650$     &  $\pm 0.086$  & B  \\
Effective temperature, $T_{\rm eff}$~[K]    &  $6429$      &  $\pm 100$      & D  \\
Surface gravity, $\log g_\star$~[cgs]       &  $4.244$     &  $\pm 0.041$    & E  \\
Projected rotation rate, $v\sin i_\star$~[km~s$^{-1}$]  &  $18.54$    &  $\pm 0.17$  & C  \\
Metallicity, [Fe/H]                          &  $-0.177$   &  $\pm 0.080$    & D  \\
Luminosity [L$_\odot$]                      &  $2.92$     &  $-0.48$, $+0.59$   & E  \\ 
Age [Gyr]                                   &  $2.82$      &  $-0.82$, $+0.58$  & E  \\ 
Distance [pc]                               &  $174$      &  $\pm 18$       & E  \\ 
& & & \\
{\it Planetary parameters:} & & & \\
$M_p$~[M$_{\rm Jup}$]                               &  $11.79$   &  $\pm 0.59$   & E  \\ 
$R_p$~[R$_{\rm Jup}$]                               &  $1.217$   &  $\pm 0.073$   & E  \\ 
Surface gravity, $\log g_p$~[cgs]                 &  $4.295$   &  $\pm 0.042$     & B  \\
Mean density, $\rho_p$~[g~cm$^{-3}$]               &  $8.1$      &  $-1.3$, $+1.7$  & E  \\
Equilibrium temperature, $T_{\rm eff}(R_\star/a)^{1/2}$~[K] &  $1710$     &  $\pm 46$       & E
\enddata

\tablecomments{ (A) Determined independently from our joint analysis
  of the photometric and radial-velocity data. (B) Functions of group
  A parameters. (C) From \jk. (D) From \jk, with enlarged error bars
  (see \S~4.1). (E) Functions of group A parameters, supplemented by
  results of the isochrone analysis (see \S~4.1) to break the
  degeneracies $M_p\propto M_\star^{2/3}$, $R_p \propto R_\star
  \propto M_\star^{1/3}$.}

\end{deluxetable}

\begin{deluxetable}{lccc}
\tabletypesize{\scriptsize}
\tablecaption{Limb-Darkening Parameters for XO-3\label{tbl:ldark}}
\tablewidth{0pt}

\tablehead{
\colhead{Bandpass} & \multicolumn{2}{c}{Tabulated Values} & \colhead{Fitted Value of} \\
\colhead{}         & \colhead{Linear Coefficient} & \colhead{Quadratic Coefficient} & \colhead{Linear Coefficient}
}

\startdata
$z$    & $0.13$ & $0.35$ & $0.11 \pm 0.07$ \\
$I$    & $0.16$ & $0.36$ & $0.06 \pm 0.15$ \\
$R$    & $0.23$ & $0.37$ & $0.16 \pm 0.14$ \\
$V$    & $0.31$ & $0.36$ & $0.47 \pm 0.14$ \\
Clear\tablenotemark{a} & \nodata & $0.33$ & $0.47 \pm 0.13$
\enddata

\tablecomments{The assumed limb-darkening law was $I_\mu/I_0 = 1 -
  a(1-\mu) - b(1-\mu)^2$.  The tabulated coefficients in Columns 2 and
  3 are based on interpolation the ATLAS tables of Claret~(2000,
  2004), for the stellar parameters $T_{\rm eff} = 6429$~K, $\log g =
  4.244$~(cgs), [Fe/H]$=-0.177$ and $v_t = 2.0$~km~s$^{-1}$. Column 4
  gives the results of fitting for the linear coefficient when the
  quadratic coefficient is fixed at the tabulated value.}

\tablenotetext{a}{This entry refers to the (unfiltered) Wise data, for
  which we used the tabulated quadratic coefficient $b=0.33$
  appropriate for the SDSS $g$ band.}

\end{deluxetable}

\end{document}